\documentclass[%
reprint,preprintnumbers,
nofootinbib,
amsmath,amssymb,
aps
prl
]{revtex4-1}
\pdfoutput=1
\usepackage{graphicx}
\usepackage[utf8]{inputenc}
\usepackage{flushend,comment}
\usepackage{dcolumn}
\usepackage{bm}
\usepackage[normalem]{ulem}
\usepackage[colorlinks = true,
            linkcolor = blue,
            urlcolor  = blue,
            citecolor = blue,
            anchorcolor = blue]{hyperref}
\usepackage{verbatim}
\usepackage{color,ulem}
\usepackage{amsmath}
\usepackage[english]{babel}

\newcommand{\beq}{\begin{eqnarray}}
\newcommand{\eeq}{\end{eqnarray}}
\def\be{\begin{equation}}
\def\ee{\end{equation}}
\def\bea{\begin{eqnarray}}
\def\eea{\end{eqnarray}}
\newcommand{\dd}{\mathrm{d}}
\definecolor{applegreen}{rgb}{0.55, 0.71, 0.0}

\begin{document}

\preprint{APCTP Pre2026 - 010}

\title{Strange-Metal Magnetotransport without a Strange-Metal Ground State}

\author{Yongjun Ahn$^{1,2,3}$}\email{yongjun.ahn@apctp.org}
\author{Matteo Baggioli$^{1,2}$}\email{b.matteo@sjtu.edu.cn}
\author{Hyun-Sik Jeong$^{3,4}$}\email{hyunsik.jeong@apctp.org}
\author{Keun-Young Kim$^{5,6}$ \vspace{1.6mm}}\email{fortoe@gist.ac.kr}

\affiliation{$^{1}$School of Physics and Astronomy, Shanghai Jiao Tong University, Shanghai 200240, China}
\affiliation{$^{2}$Wilczek Quantum Center, School of Physics and Astronomy, Shanghai Jiao Tong University, Shanghai 200240, China}
\affiliation{$^{3}$Asia Pacific Center for Theoretical Physics, Pohang 37673, Korea}
\affiliation{$^4$Department of Physics, Pohang University of Science and Technology, Pohang 37673, Korea}
\affiliation{$^5$Department of Physics and Photon Science, Gwangju Institute of Science and Technology, 123 Cheomdan-gwagiro, Gwangju 61005, Korea}
\affiliation{$^6$Research Center for Photon Science Technology, Gwangju Institute of Science and Technology, 123 Cheomdan-gwagiro, Gwangju 61005, Korea}

\begin{abstract}
Strange metals display a linear-in-temperature resistivity, $\rho_{xx}\sim T$, and a quadratic inverse Hall angle, $\cot\Theta_H\sim T^2$. These seemingly universal laws are often attributed to quantum criticality, yet no controlled theory realizes both at an asymptotic zero-temperature fixed point, and experimentally they persist only over finite temperature windows. Here, we show that these properties can arise from an intermediate critical regime that does not govern the ground state. Using a minimal class of holographic models, we construct a fully backreacted realization of both scalings, which appear within a finite temperature window controlled by an emergent intermediate scaling geometry. The window is bounded by relevant deformations of the corresponding fixed point: momentum relaxation sets the upper edge, while the magnetic field $B$ closes it from below at $T_{\min}\sim B^{2/3}$, up to logarithmic corrections. At $T_{\min}$, strange-metal transport, the scaling geometry, and critical thermodynamics break down together. Our results point to intermediate criticality as a distinct route to strange-metal transport, and suggest that it can be identified by the simultaneous breakdown of transport and thermodynamic scaling under a magnetic field.
\end{abstract}

\maketitle
\begingroup
\renewcommand\thefootnote{}
\footnotetext{Authors are listed in alphabetical order.}
\endgroup

\textbf{Introduction} -- Strange metals are metallic states whose transport properties depart strongly from conventional quasiparticle expectations, displaying unconventional temperature scalings over extended regimes \cite{Phillips:2022nxs}. Among their most prominent signatures are a linear-in-temperature resistivity and a quadratic inverse Hall angle,
\begin{equation}
\rho_{xx}\sim T,
\qquad
\cot\Theta_H\sim T^2 ,
\label{eq:SM}
\end{equation}
observed over broad temperature ranges in cuprates, heavy-fermion compounds,
and other correlated systems
\cite{PhysRevLett.59.1337,PhysRevLett.69.2975,PhysRevLett.67.2088,
doi:10.1126/science.1165015,Bruin:2013qlp,
custers2010evidence,paschen2004hall,
doi:10.1126/science.1262054,custers2003break}.
Their coexistence is particularly striking because standard Drude transport
ties longitudinal and Hall transport to a single relaxation rate, thereby
predicting the same temperature dependence
\cite{PhysRevLett.67.2092,Mizutani_2001}.

While such power laws are often associated with an underlying
quantum critical point, experiments routinely observe strange-metal behavior
over finite temperature windows before the system enters another low-temperature phase, such as a superconducting state. This raises a
fundamental question: \emph{must strange-metal transport originate from an
asymptotic zero-temperature fixed point, or can it instead emerge from an
intermediate critical regime along the renormalization-group (RG) flow?}
The latter scenario would imply that observed scaling exponents need not
uniquely encode the nature of the ultimate ground state.

Holography provides a controlled theoretical framework in which both transport observables and the underlying RG flow can be tracked explicitly \cite{hartnoll2018,Zaanen:2015oix}. In holographic systems without Galilean
invariance, the longitudinal DC conductivity naturally decomposes into two
distinct channels: a charge-conjugation-symmetric contribution
$\sigma_{\mathrm{ccs}}$ and a dissipative Drude-like contribution
$\sigma_{\mathrm{diss}}$ associated with momentum relaxation. Meanwhile,
in the weak-magnetic-field limit the Hall angle is controlled by the
dissipative channel,
$\tan\Theta_H\sim (B/q)~\sigma_{\mathrm{diss}}$ at finite charge density $q$
\cite{Blake:2014yla}. Hence, when
$\sigma_{\mathrm{ccs}}\sim T^{-1}$ and
$\sigma_{\mathrm{diss}}\sim T^{-2}$, provided $\sigma_{\mathrm{ccs}}\gg \sigma_{\mathrm{diss}}$, the longitudinal resistivity can be
linear in temperature while the inverse Hall angle remains quadratic,
providing a natural route to Eq.~\eqref{eq:SM} \cite{Blake:2014yla}.

\begin{figure*}[ht]
    \includegraphics[width=\linewidth]{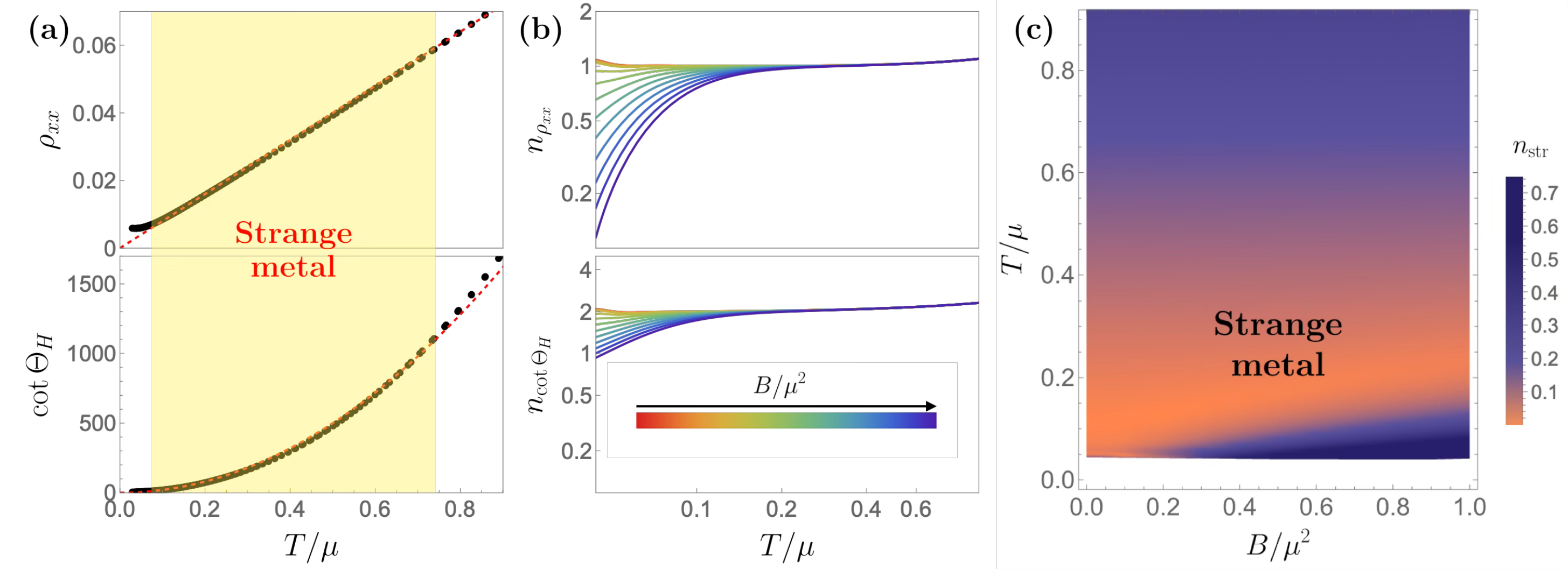}
    \caption{\textbf{Strange-metal magnetotransport in an intermediate temperature window.} \textbf{(a)} Longitudinal resistivity $\rho_{xx}$ and inverse Hall angle $\cot\Theta_H$ as functions of $T/\mu$ at $B/\mu^2=1$ and $k/\mu=80$. Red dashed lines are fits $\rho_{xx}\propto T$ and $\cot\Theta_H\propto T^2$. The yellow region marks the strange-metal window. \textbf{(b)} Local transport exponents $n_{\rho_{xx}}$ and $n_{\cot\Theta_H}$ as functions of $T/\mu$ at $k/\mu=80$, for $B/\mu^2\in[0,1]$ increasing from red to blue. Strange-metal transport corresponds to $(n_{\rho_{xx}},n_{\cot\Theta_H})\simeq(1,2)$. \textbf{(c)} Deviation from ideal strange-metal transport, $n_{\mathrm{str}}$, in the $(B/\mu^2,\,T/\mu)$ plane at $k/\mu=80$. The orange region, $n_{\mathrm{str}}\ll 1$, identifies the strange-metal window.}
    \label{fig:1}
\end{figure*}

Importantly, this mechanism was already recognized not to persist
asymptotically down to $T\to0$. The $T$-linear resistivity emerges only above
a model-dependent microscopic crossover scale $W$, such that the
strange-metal regime requires $T\gg W$ \cite{Blake:2014yla}. Since the Hall
angle is defined only at finite magnetic field, $B$ cannot be treated as a
spectator: it is a necessary ingredient of strange-metal magnetotransport,
and its role in affecting the strange-metal regime has, to our knowledge, not
been addressed in a fully backreacted setting. 

Moreover, scaling analyses showed that the simultaneous realization of Eq.~\eqref{eq:SM} is
forbidden in the asymptotic low-temperature limit of broad classes of
holographic scaling geometries \cite{Amoretti:2016cad,Ahn:2023ciq}.
Accordingly, to the best of our knowledge, a controlled realization of both
scalings from a single asymptotic zero-temperature fixed point remains absent,
both within holography and in other theoretical approaches. This motivates a
different possibility: \textit{the critical regime responsible for
strange-metal transport may not be the deep infrared, but an intermediate
regime along the RG flow.} Very recently, a superlinear Hall angle coexisting
with linear-in-temperature resistivity was reported in a microscopic
Yukawa-SYK model~\cite{32ts-qh8d}, within an extended crossover regime above its
low-temperature marginal Fermi liquid ground state, providing preliminary
support for this scenario. Here, we realize this scenario explicitly in a minimal, fully backreacted
holographic model, and trace the strange-metal window to an emergent
intermediate scale covariant unstable fixed point whose relevant deformations, momentum relaxation
and the magnetic field, set its upper and lower edges.

\textbf{Holographic setup} -- We consider the 4D Einstein-Maxwell-Dilaton-Axion (EMDA) holographic model~\cite{Gouteraux:2014hca},
\begin{equation}
\label{eq:action}
\begin{aligned}
S=\int d^4x\,\sqrt{-g}\Big[
&R-\frac12(\partial\phi)^2+V(\phi)
-\frac{Z(\phi)}{4}F_{MN}F^{MN} \\
&-\frac{Y(\phi)}{2}\sum_{i=1}^{2}(\partial\chi_i)^2
\Big],
\end{aligned}
\end{equation}
where $\phi$, $F=dA$, and $\chi_i$ denote the dilaton, gauge field strength,
and axions, respectively. We make the following ansatz
\begin{align}
\dd s^2 &= -D(r)\,\dd t^2 + \mathcal{B}(r)\,\dd r^2
+ C(r)\,\dd x_i^2 \,, \qquad \phi = \phi(r),
\nonumber\\
A&=A_t(r)\,\dd t-\frac{B}{2}(y\,\dd x-x\,\dd y),
\qquad
\chi_i=kx_i \,.\label{eq:ansatz:gen}
\end{align}
where $B$ is the magnetic field and the parameter $k$ controls the rate of momentum relaxation \cite{Andrade:2013gsa,Baggioli:2021xuv}.
The temperature of the dual system is given by $T=\frac{1}{4\pi}\frac{|D'_h|}{\sqrt{D_h\mathcal{B}_h}}$ where the subscript $_h$ denotes that the quantity is evaluated at the black hole horizon. 
The charge density $q$ is defined as $q=\frac{C Z A_t'}{\sqrt{D\mathcal{B}}}$. 
The DC conductivities of the dual system are given by~\cite{Donos:2014uba, Blake:2014yla}
\begin{align}\label{conductivities:gen}
\begin{split}
\sigma_{xx} &= \frac{k^2 \, C_h \left( q^2 +B^2 Z_h^2 + k^2\,C_h Z_h \right)}{B^2 q^2 + \left( B^2 Z_h + k^2 \, C_h \right)^2}  \,,\\
\sigma_{xy} &= \frac{B q \left( q^2 +B^2 Z_h^2 + 2 k^2 \, C_h Z_h \right)}{B^2 q^2 + \left( B^2 Z_h + k^2 C_h \right)^2}  \,.
\end{split} 
\end{align}

Our key transport coefficients are the longitudinal resistivity and the Hall angle, which can be computed from the conductivities as $ \rho_{xx} = \frac{\sigma_{xx}}{\sigma_{xx}^2+\sigma_{xy}^2}$ and $\cot \Theta_H\equiv\frac{\sigma_{xx}}{\sigma_{xy}}$.
At vanishing magnetic field, the charge-conjugation symmetric contribution and Drude-like term can be read off as $\sigma_{\text{ccs}}\equiv Z_h$, $\sigma_{\text{diss}} \equiv \frac{q^2}{k^2C_h}$.

\textbf{Strange-metal transport from intermediate criticality} --
We now identify the regime responsible for the characteristic strange-metal
transport laws. At vanishing magnetic field, a running dilaton with
$Y(\phi)=1$ admits semi-locally critical, or $\eta$-geometry, solutions
\cite{Gouteraux:2014hca},
\begin{align}
\dd s^2 &=
r^{-\eta}
\left[
\frac{1}{r^2}
\left(
\frac{L^2}{f(r)}\dd r^2-f(r)\dd t^2
\right)
+\widetilde L^{\,2}\dd x_i^2
\right],
\nonumber\\
A_t &= Q r^{\zeta-1},
\qquad
\phi=\kappa\log r \,,\qquad\kappa^2=\eta(\eta+2)\,,\nonumber\\
\qquad L^2&=\frac{(\eta+1)(\eta+2)}{V_0}\,,\qquad \widetilde{L}^2=\frac{k^2(\eta+2)}{2V_0} \,,
\label{ansatz:etageom}
\end{align}
with $f(r)=1-(r/r_h)^{\eta+1}$ and $r_h\sim T^{-1}$, where $V_0$ is defined by $V(\phi)\simeq V_0\,e^{\delta\phi}$ at large $\phi$ ($V_0=3$ for the potential used below), and $\zeta=-1$ for $\eta=2$. On the branch considered here, the gauge field enters as a perturbation of the neutral scaling solution (see Supplemental Material (SM)~\cite{SM}).

We focus on the branch with relevant electric current and marginally relevant momentum relaxation. For the strange-metal solution, $\eta=2$, and the two conductivity channels scale as $\sigma_{\mathrm{ccs}}\sim T^{-1}$ and $\sigma_{\mathrm{diss}}\sim T^{-2}$. In the strong momentum-relaxation regime, $\sigma_{\mathrm{ccs}}\gg\sigma_{\mathrm{diss}}$, so that the former dominates the longitudinal conductivity, while the latter controls the Hall response at weak magnetic field. It then follows directly that $\rho_{xx}\sim T$ and $\cot\Theta_H\sim T^2$. This realizes the mechanism of Ref.~\cite{Blake:2014yla} in a fully backreacted solution. 

However, both the charge density and the magnetic field are relevant perturbations of this fixed point. At $B=0$, the charge deformation grows toward the IR as $X_q\propto q^2 r/k^4$ (see SM~\cite{SM}), and at the horizon it equals $X_q=\sigma_{\mathrm{diss}}/\sigma_{\mathrm{ccs}}=W/T$: the Blake-Donos crossover at $T\sim W$ is therefore also the scale at which the charge deforms the $\eta=2$ geometry. At finite field, $B$ provides a second relevant perturbation, with $X_B\propto B^2 r^3/k^4$. The $\eta=2$ solution thus never describes the asymptotic low-temperature state, but governs an intermediate scaling regime whose lower extent is set by $\max(W,T_{\min})$, with $T_{\min}$ controlled by $B$. In the strong momentum-relaxation regime considered here, $T_{\min}\gg W$, so that the magnetic field sets the lower edge.

The linear mode analysis around the $\eta=2$ solution is given in the SM~\cite{SM}.
In the \textit{End Matter} we also present a hyperscaling-violating realization with
$Y(\phi)\neq1$, showing that the mechanism extends beyond the specific model
considered here.

For the numerical construction, we choose
$V(\phi)=6\cosh(\phi/\sqrt{2})-\phi^2/2$ and
$Z(\phi)=e^{\phi/(2\sqrt{2})}$, whose large-$\phi$ behavior realizes the
exponents required for strange-metal transport. We numerically construct the
corresponding asymptotically $\mathrm{AdS}_4$ solutions at finite charge
density, momentum relaxation, and magnetic field, and evaluate the transport
coefficients using the horizon formulae above. Details of the numerical
setup are given in the SM~\cite{SM}.

Fig.~\ref{fig:1}(a) provides a direct demonstration of the strange-metal
transport window (yellow shaded region), where $\rho_{xx}\propto T$ and
$\cot\Theta_H\propto T^2$ simultaneously for fixed momentum relaxation and
magnetic field.

To quantify this intermediate regime, we define the local transport exponents
$n_{\rho_{xx}}\equiv d\log\rho_{xx}/d\log T$ and
$n_{\cot\Theta_H}\equiv d\log\cot\Theta_H/d\log T$.
Figure~\ref{fig:1}(b) shows these quantities as functions of temperature for
different values of the dimensionless magnetic field $B/\mu^2$ at fixed
momentum relaxation. A clear regime with
$(n_{\rho_{xx}},n_{\cot\Theta_H})\simeq(1,2)$ emerges, whose lower-temperature
boundary moves to higher temperatures as the magnetic field is increased.

Finally, we characterize the full strange-metal region through the combined
deviation
$n_{\mathrm{str}}\equiv
\sqrt{(n_{\rho_{xx}}-1)^2+(n_{\cot\Theta_H}-2)^2}$. In our notation, $n_{\mathrm{str}} \ll 1$ identifies strange-metal magnetotransport. Figure~\ref{fig:1}(c) shows $n_{\mathrm{str}}$ in the $(B/\mu^2,\,T/\mu)$ plane, providing a direct map of the strange-metal
transport regime and its field-dependent lower boundary.

\begin{figure}[t]
 \includegraphics[width=0.9\linewidth]{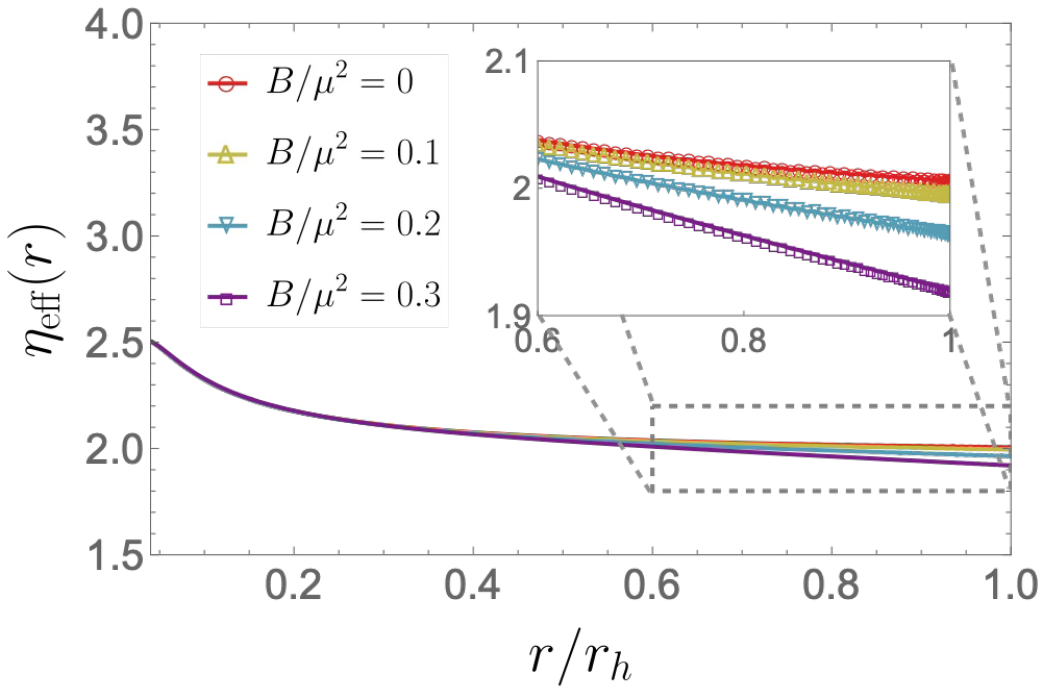}
    \caption{\textbf{Emergence of the intermediate scaling geometry.} Effective scaling exponent $\eta_{\mathrm{eff}}(r)$ as a function of the rescaled bulk radial coordinate $r/r_h$, from the AdS$_4$ boundary ($r=0$) to the horizon ($r=r_h$), at fixed $k/\mu=80$ and $T/\mu\approx\,0.45$. Colors correspond to $B/\mu^2=0,\,0.1,\,0.2,\,0.3$. Away from the near-boundary region, $\eta_{\mathrm{eff}}$ approaches 2, signaling the $\eta=2$ geometry responsible for strange-metal transport. The inset shows that increasing $B$ drives $\eta_{\mathrm{eff}}$ away from $2$ near the horizon, as expected for a relevant deformation of the intermediate fixed point.}
    \label{fig:2}
\end{figure}
\textbf{Intermediate scaling geometry} -- In the framework of the renormalization group, the intermediate strange metal region corresponds to an intermediate scaling regime that is governed by an unstable fixed point that does not correspond to the ultimate infrared (IR) ground state. Within the holographic framework, the RG flow is encoded in the dynamics along the radial extra-dimension \cite{Baggioli2019HolographicRG}. Hence, an intermediate scaling regime must coincide directly with the emergence of an intermediate geometry with scaling properties, that is exactly the $\eta=2$ geometry discussed above.

To show this, we construct a local scaling exponent $\eta_{\mathrm{eff}}(r)$ that depends on the radial coordinate $r$. The geometrical definition of this parameter is given by $\eta_{\rm eff}(r)\equiv 2\,(\log C)'/[\log(D\mathcal{B}/C'^{2})]'$. A constant $\eta_{\mathrm{eff}}$ indicates the existence of a scaling $\eta$ geometry characterized by such exponent. The UV boundary is located in this choice of coordinates at $r=0$, while the IR thermal horizon at $r=r_h$. This means that moving from $r=0$ to $r=r_h$ is equivalent to following the RG flow from the UV fixed point down to the infrared scale set by the temperature. The UV fixed point is dual by construction to our AdS$_4$ geometry and corresponds to a relativistic three-dimensional conformal field theory (CFT$_3$) that is deformed by various operators including one dual to the axion field and responsible for the explicit breaking of translational symmetry.

Fig.~\ref{fig:2} presents a plot of $\eta_{\mathrm{eff}}(r)$ as a function of $r$ for fixed momentum relaxation rate and varying strength of the dimensionless external magnetic field $B/\mu^2$. In a large intermediate range of values, $\eta_{\mathrm{eff}}(r) \approx 2$ and constant, indicating the emergence of the intermediate scaling regime. By increasing the magnetic field, we see increasing deviations in the IR towards $r \rightarrow r_h$ indicating that the extent of this regime into the IR becomes smaller and smaller. This is consistent with the idea, verified explicitly in the SM~\cite{SM}, that the magnetic field acts as a relevant deformation around the $\eta=2$ fixed point pushing the system towards another IR ground state that does not exhibit strange metal properties.

\textbf{Edges of the strange-metal window} -- Leveraging the geometrization of the RG flow and the theoretical control over the holographic model, we can analytically derive the edges of the intermediate strange-metal window.
The magnetic field is a relevant deformation of the $\eta=2$ scaling solution and therefore eventually drives the RG flow away from this intermediate fixed point toward the IR. Linearizing the bulk equations around the $\eta=2$ solution (see SM~\cite{SM}), one finds a magnetic deformation whose scaling dimension directly implies
\begin{equation}
\label{eq:Tmin}
T_{\min}\sim k\left(\frac{B}{k^2}\right)^{2/3}\log^{1/3}\left(\frac{k^4}{B^2}\right),
\end{equation}
which reduces to a power-law scaling $T_{\min}\sim B^{2/3}k^{-1/3}$ up to a slowly varying logarithm. The logarithm is not accidental: the magnetic deformation resonates with the thermal mode of the $\eta=2$ geometry precisely when $\gamma\kappa=1$, which is the condition for $\rho_{xx}\sim T$ (see SM~\cite{SM}). The logarithmic correction is therefore a direct consequence of linear-in-$T$ resistivity.

The same scaling can be understood heuristically from the competition between the terms supporting the $\eta=2$ background and the magnetic Maxwell contribution. Evaluated at the horizon, they scale respectively as
\begin{equation}
\frac{k^2}{C_h}\sim k^2T^{-2},\qquad Z(\phi)F_B^2\sim \frac{B^2Z_h}{C_h^2}\sim B^2kT^{-5},
\end{equation}
where we used that, for marginally relevant momentum relaxation, the IR scale is set by $k$, so that $C_h\propto T^2$ and $Z_h\propto k/T$. The magnetic deformation becomes important when these two contributions are comparable, recovering $T_{\mathrm{min}}\sim B^{2/3}k^{-1/3}$.

Up to logarithmic corrections, the same scale can be directly derived from the holographic DC conductivities. Within the intermediate $\eta=2$ scaling regime, and at small magnetic field, the transport coefficients in Eq.~\eqref{conductivities:gen} take the form
\begin{equation}\label{eq:resgen}
\begin{split}
\rho_{xx}&\sim T\frac{1+W/T+W_{\Theta_H}^3/T^3}{\left(1+W/T\right)^2+WW_{\Theta_H}^3/T^4}\,,\\ \cot\Theta_H&\sim \frac{T^2}{\sqrt{WW_{\Theta_H}^3}}\frac{1+W/T+W_{\Theta_H}^3/T^3}{2+W/T+W_{\Theta_H}^3/T^3}\,,
\end{split}
\end{equation}
where the horizon scaling relations, $Z_h\propto k/T$ and $C_h\propto T^2$, imply $W\sim q^2/k^3$ and $W_{\Theta_H}^3\sim B^2/k$. Strange-metal transport therefore requires $T\gg \max\!\left(W, W_{\Theta_H}\right)$.

At $B=0$, the scale $W$ provides the low-temperature crossover originally identified by Blake and Donos \cite{Blake:2014yla}: below $T\sim W$, the momentum-relaxing contribution becomes important and the linear resistivity is lost. Since $W/T=X_q$ at the horizon, this is also where the charge deformation backreacts on the intermediate geometry.

At finite magnetic field, however, Eq.~\eqref{eq:resgen} contains the additional scale $W_{\Theta_H}$. In the strong momentum-relaxation regime relevant here, $W\propto k^{-3}$ is parametrically more suppressed than $W_{\Theta_H}\propto k^{-1/3}$. Consequently, over the regime explored numerically one has $W_{\Theta_H}>W$, and the first departure from strange-metal transport upon lowering the temperature is controlled by the magnetic scale, $T_{\mathrm{min}}\sim W_{\Theta_H}\sim B^{2/3}k^{-1/3}$, exactly the scale at which the intermediate geometry breaks down.

This coincidence of the transport and geometric scales, in both $B$ and $k$, has a concrete physical origin. To leading order in $W/T$, the magnetic field enters the DC conductivities only through the combination $X_B\equiv B^2Z_h/(k^2C_h)=W_{\Theta_H}^3/T^3\sim B^2/(kT^3)$ , i.e., the strength of the magnetic deformation at the black brane horizon. The same ratio measures the magnetic Maxwell stress, $Z F_B^2\sim B^2Z_h/C_h^2$, relative to the axion stress $k^2/C_h$ that supports the $\eta=2$ background. The magnetic field thus starts to affect transport precisely when it starts to backreact on the geometry, $X_B\sim1$, and strange-metal transport ends at the same place as the intermediate critical regime. The absence of logarithmic corrections in the transport estimate simply reflects the pure power-law approximation used for $Z_h$ and $C_h$.

Finally, the upper boundary of the strange-metal window is set by momentum relaxation, $T_{\max}\sim k$. Near the AdS$_4$ fixed point, $T$ and $k$ both have scaling dimension one, and for $k/\mu \gg 1$ and $B \ll k^2$, $k$ is the dominant UV scale, so $T_{\max}\sim k$.

\begin{figure}[ht]
\centering
\includegraphics[width=0.99\linewidth]{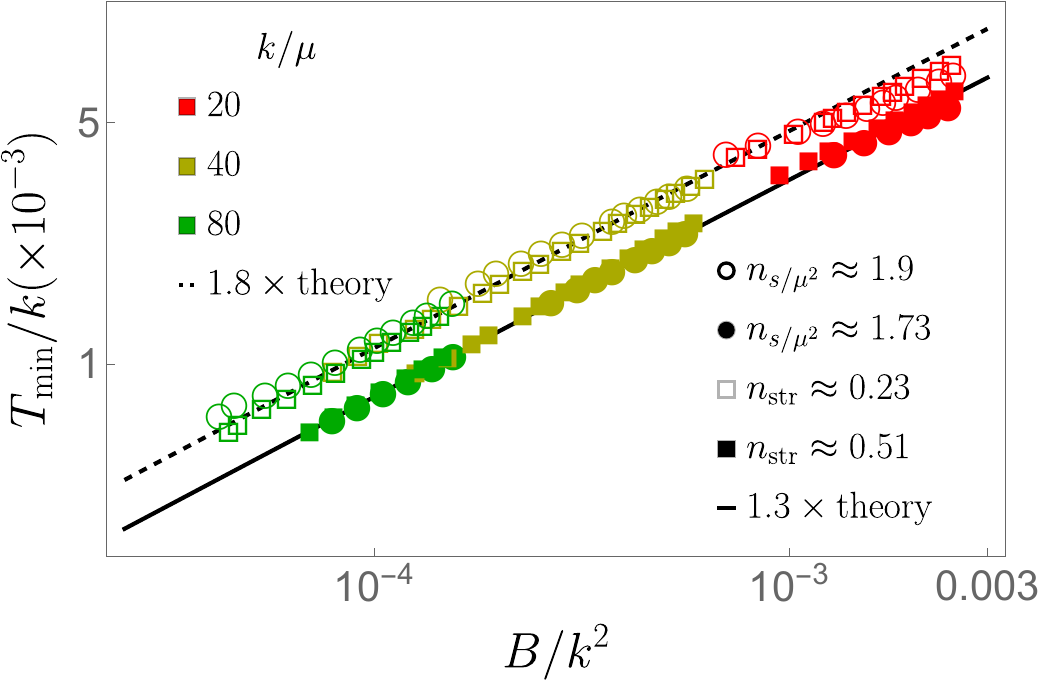}
\caption{\textbf{Lower edge of the strange-metal window.} $T_{\min}/k$ as a function of $B/k^2$ on logarithmic scales, for $k/\mu=20$ (red), $40$ (yellow), and $80$ (green). Squares mark the breakdown of strange-metal transport, defined by $n_{\mathrm{str}}\approx 0.23$ (open) and $0.51$ (filled). Circles mark the departure of the entropy exponent $n_s$ from its critical value $n_s=2$, defined by $n_s\approx1.9$ (open) and $1.73$ (filled). The thresholds form matched pairs: within each pair, transport and entropy data coincide and, for all values of $k$, collapse onto the analytical prediction, Eq.~\eqref{eq:Tmin}, with a single overall constant (dotted line: open symbols; solid line: filled symbols). Changing the threshold changes only this constant, not the scaling.}
\label{fig:3}
\end{figure}

We test our theoretical predictions numerically in Fig.~\ref{fig:3}, which shows $T_{\min}/k$ as a function of $B/k^2$ for $k/\mu=20,40,80$. We extract $T_{\min}$ independently from two criteria: the breakdown of strange-metal transport, defined by $n_{\mathrm{str}}\approx 0.23$, and the departure of the entropy exponent $n_s\equiv d\log s/d\log T$ from its critical value $n_s=2$ in the $\eta=2$ geometry, defined by $n_s\approx 1.9$. For all values of $k$, the two data sets follow the same scaling and collapse onto the analytical prediction, Eq.~\eqref{eq:Tmin}, up to an overall constant. The collapse is insensitive to the precise choice of thresholds, as directly verified using two different cutoffs (see Fig.~\ref{fig:3} and caption therein). The scaling, its $k$-dependence, and the IR scale entering Eq.~\eqref{eq:Tmin} are fixed by the analysis; the remaining $\mathcal{O}(1)$ constant depends on the threshold chosen to define the breakdown. The residual drift of this constant with $k$ reflects subleading corrections in $\mu/k$: the IR scale $\Lambda_{\mathrm{IR}}/k$ (see \textit{End Matter}) depends weakly on $k/\mu$, and the charge deformation $X_q\propto q^2r/k^4$ adds to the magnetic one at finite $\mu/k$. Both corrections vanish for $k/\mu\to\infty$, where the collapse is expected to become exact. 

This confirms both the scaling $T_{\min}\sim B^{2/3}k^{-1/3}$, including its dependence on $k$, and the central result of this work: up to a threshold-dependent constant that does not depend on $B$ or $k$, strange-metal transport and critical thermodynamics end at the same temperature, set by the magnetic deformation of the intermediate fixed point. Together with the upper edge $T_{\max}\sim k$, verified in \textit{End Matter}, this fixes both edges of the strange-metal window analytically. Combining both edges, the strange-metal window spans $T_{\max}/T_{\min}\sim (k^2/B)^{2/3}$, up to logarithmic corrections, and is therefore parametrically wide for strong momentum relaxation and weak magnetic field.

\textbf{Discussion} -- Our results show that strange-metal magnetotransport need not originate from the ultimate IR fixed point. It can instead be inherited from an intermediate critical regime, whose edges are set by the relevant deformations of the RG flow. A direct consequence, which we verified explicitly, is that the field-induced end of strange-metal transport coincides with the crossover in thermodynamic quantities, since both are controlled by the same magnetic deformation of the intermediate geometry. More broadly, this implies that the observed transport exponents need not encode the nature of the ultimate ground state.

Testing whether the same coincidence holds in microscopic strange-metal models at finite field~\cite{32ts-qh8d,Patel2023Universal} would be a natural next step. Experimentally, simultaneous transport and thermodynamic measurements under a magnetic field, such as resistivity and specific heat across the field-induced end of $T$-linear resistivity, could distinguish an intermediate critical regime from a quantum critical ground state.

\textbf{Acknowledgments} -- MB thanks K.~Skenderis and N.~Obers for useful discussions. MB acknowledges the support of the Foreign Young Scholars Research Fund Project (Grant No. 22Z033100604) and the sponsorship from the Yangyang Development Fund.
HSJ was supported by an appointment to the JRG Program at the APCTP through the Science and Technology Promotion Fund and Lottery Fund of the Korean Government. HSJ was also supported by the Korean Local Governments -- Gyeongsangbuk-do Province and Pohang City.
KYK was supported by the National Research Foundation of Korea (NRF) grant funded by the Korea government (MSIT) (RS-2026-25477372, RS-2025-02311201, RS-2025-02307394, RS-2024-00445164), and by the Creation of the Quantum Information Science R$\&$D Ecosystem (Grant No. 2022M3H3A106307411, RS-2023-NR068116). KYK was also supported by GIST research fund [Future leading Specialized Research Project, 2026 and the ANCHOR program through the Gwangju ANCHOR Center, funded by the Ministry of Education (MOE) and the Jeonnam-Gwangju Special Metropolitan City, Republic of Korea, (2026-ANCHOR-05-001)].

\bibliographystyle{apsrev4-1}
%

\section*{End Matter}\label{end}
\textit{Appendix A: General hyperscaling-violating family} --  In the main text, we focused on the simple case of a trivial axion-dilaton coupling, $Y(\phi)=1$. The same mechanism, however, extends to the more general class $Y(\phi)\sim e^{\lambda\phi}$, for which the infrared geometry is hyperscaling violating rather than semi-locally critical. In this case, the temperature scalings of the longitudinal resistivity and Hall angle cotangent are given by~\cite{Amoretti:2016cad} [after substituting $k^2 \rightarrow k^2 Y(\phi_h)$ in Eq. \eqref{conductivities:gen}]
\begin{align}
&\rho_{xx}\sim
T^{\frac{2\gamma(\delta-\lambda)}
{1-\delta^2+\lambda^2}}
=
T^{\frac{2-\widetilde{\zeta}}{z}},\\
&\cot\Theta_H\sim
T^{\frac{2(\delta-2\lambda)(\delta-\lambda)}
{1-\delta^2+\lambda^2}}
=
T^{\frac{4-\theta}{z}}.
\end{align}
Here, $\widetilde{\zeta}$ denotes the conduction exponent associated with the gauge field. Requiring simultaneously $\rho_{xx}\sim T$ and $\cot\Theta_H\sim T^2$ gives $z=2-\widetilde{\zeta}$, $\theta=2\widetilde{\zeta}$. These two conditions leave one exponent free and therefore define a continuous family of hyperscaling-violating geometries exhibiting the same strange-metal transport laws. The $\eta=2$ geometry studied in the main text is recovered in the semi-local limit $\widetilde{\zeta}\rightarrow-\infty$, for which $\eta\equiv-\frac{\theta}{z}=2$, $\zeta\equiv\frac{\widetilde{\zeta}}{z}=-1$.

\begin{figure}[ht]
\centering
\includegraphics[width=0.99\linewidth]{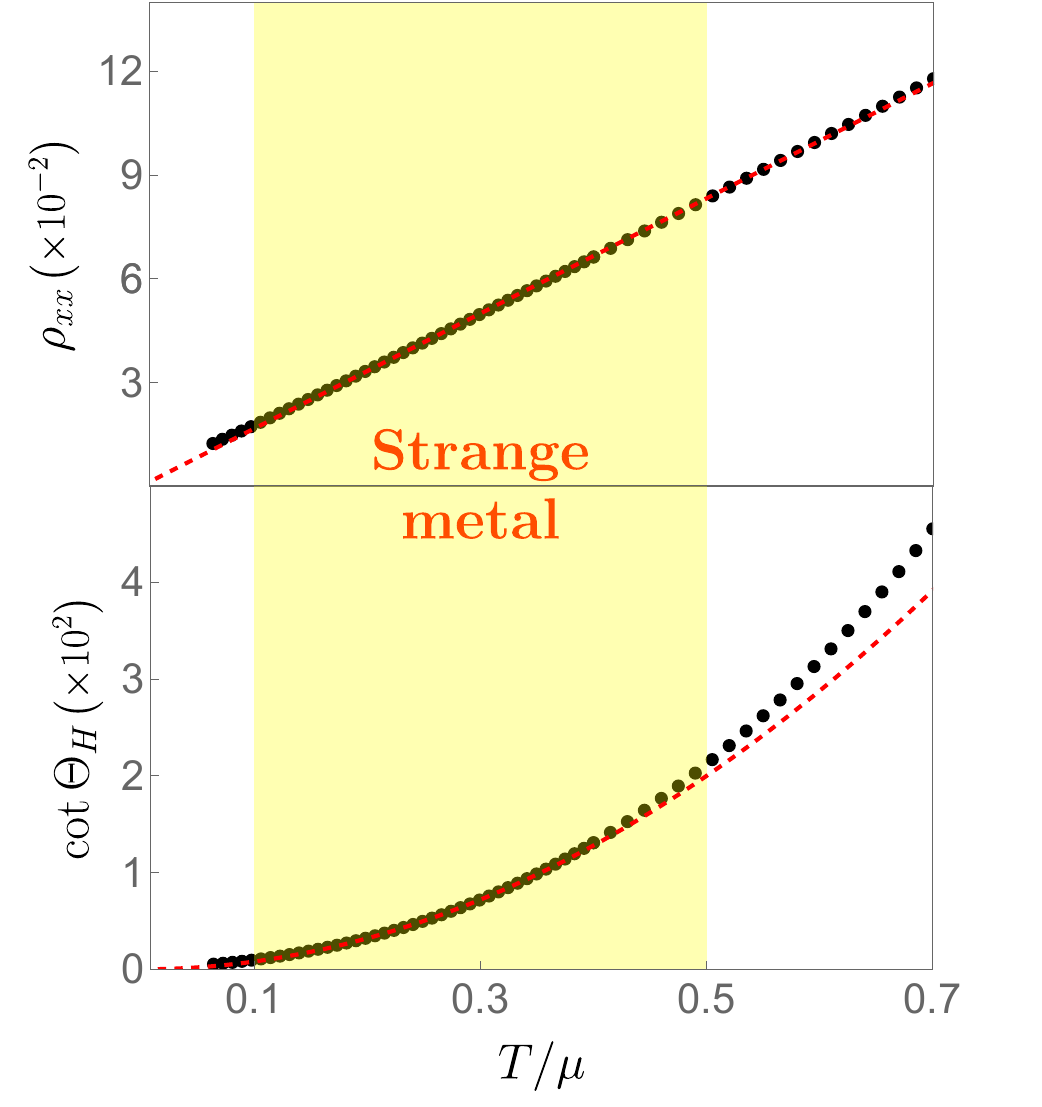}
\caption{\textbf{Strange-metal magnetotransport from hyperscaling-violating intermediate scaling solutions.} Temperature dependence of the longitudinal resistivity and Hall angle cotangent for a finite axion-dilaton coupling, $\lambda\neq0$, evaluated at $(z,\theta,\widetilde{\zeta})=(2.7,-1.4,-0.7)$, equivalently $(\gamma,\delta,\lambda)=(0.67,0.34,-0.50)$, with $B/\mu^2=0.1$ and $k/\mu=20$. Red dashed lines are fits $\rho_{xx}\propto T$ and $\cot\Theta_H\propto T^2$. The yellow region marks the strange-metal window. }
\label{fig:smfinitez}
\end{figure}

The presence of this free parameter shows that the mechanism is not tied to a single scaling geometry. While the temperature dependence of the longitudinal resistivity and Hall angle remains fixed, other physical observables can depend on the remaining exponent. This additional freedom may therefore allow the same framework to accommodate further properties of strange metals beyond the two transport scalings considered here. Figure~\ref{fig:smfinitez} provides a direct numerical check that the strange-metal transport laws survive within this more general hyperscaling-violating family. 

\begin{figure}[t]
    \includegraphics[width=\linewidth]{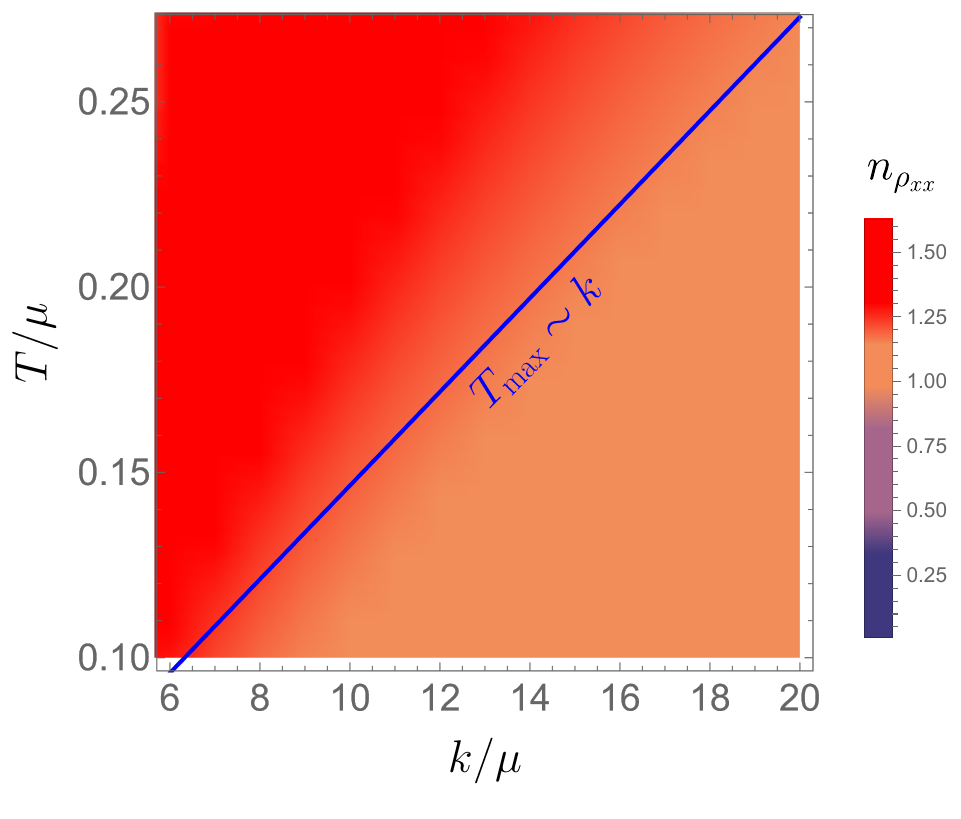}
    \caption{\textbf{The upper edge of the strange-metal region.} Local resistivity exponent $n_{\rho_{xx}}$ in the $(k/\mu,\,T/\mu)$ plane at fixed $B/\mu^2=0$. The orange region, $n_{\rho_{xx}}\simeq1$, marks linear-in-$T$ resistivity. Its upper edge follows the predicted scaling $T_{\max}\sim k$ (blue solid line).}
    \label{fig:Tmax}
\end{figure}

\textit{Appendix B: Momentum relaxation sets the upper cutoff} -- Figure~\ref{fig:Tmax} maps the local resistivity exponent $n_{\rho_{xx}}$ in the $(k/\mu,\,T/\mu)$ plane at fixed $B=0$. Increasing $k$ shifts the $n_{\rho_{xx}}\simeq1$ region to higher temperatures, so stronger momentum relaxation extends the strange-metal window toward the UV. Its upper edge follows $T_{\max}\sim k$ (blue solid line), as expected from the scaling argument in the main text.

\textit{Appendix C: Magnetic deformation and $T_{\min}$} --
For the couplings used in the main text, the large-$\phi$ exponents are $\delta=1/\sqrt{2}$ and $\gamma=1/(2\sqrt{2})$. Linearizing around the $\eta=2$ solution, a weak magnetic field deforms the IR geometry as
\begin{equation}
\begin{split}
D(r)&\sim \frac{1}{r^4}\left[1 - \frac{9B^2}{8k^4}r^3  \left(1 - 2 \log(r)\right)\right]\,,\\
\mathcal{B}(r)&\sim \frac{4}{r^4}\frac{1}{1 + \frac{9B^2}{8k^4}r^3(1 + 2 \log(r))}\,,\\
C(r)&\sim \frac{2k^2}{3r^2}\,,\\
\phi(r)&\sim 2 \sqrt{2} \log(r) + \frac{9B^2}{8\sqrt{2}k^4}r^3\,.
\end{split}
\end{equation}
The correction grows toward the IR (large $r$), consistent with $B$ being a relevant deformation, and is negligible for $r\ll r_{\max}$, with
\begin{equation}
\label{eq:rscmax}
r_{\max}\sim\left[\frac{B^2}{k^4}\,\mathcal{W}_0\left(\frac{64k^4}{3B^2}\right)\right]^{-1/3},
\end{equation}
where $\mathcal{W}_0$ is the principal branch of the Lambert function, $\mathcal{W}_0(x)e^{\mathcal{W}_0(x)}=x$. Since $r_h = 3 \Lambda_{\mathrm{IR}}/(8\pi T)$ with $\Lambda_{\mathrm{IR}} \propto k$, where $\Lambda_{\mathrm{IR}}$ is the IR energy scale fixing the normalization of the IR time coordinate, $D(r)\simeq\Lambda_{\mathrm{IR}}^2\,r^{-4}$, relative to the boundary time (set to one in the expressions above), the $\eta=2$ regime ends at
\begin{equation}
\label{eq:Tminsc}
T_{\min}\sim k\left[\frac{B^2}{k^4}\,\mathcal{W}_0\!\left(\frac{64k^4}{3B^2}\right)\right]^{1/3}\simeq k\left(\frac{B}{k^2}\right)^{2/3}\log^{1/3}\!\left(\frac{k^4}{B^2}\right),
\end{equation}
which reproduces the main text result in Eq.~\eqref{eq:Tmin}.

\clearpage
\onecolumngrid

\section*{Supplemental Material}
In this Supplemental Material, we provide more details about the holographic models and the numerical methods used. Moreover, we provide the detailed mode analysis around the intermediate scaling geometry.
\setcounter{equation}{0}
\setcounter{figure}{0}
\setcounter{table}{0}
\setcounter{section}{0}

\renewcommand{\theequation}{S\arabic{equation}}
\renewcommand{\thefigure}{S\arabic{figure}}
\renewcommand{\thetable}{S\arabic{table}}
\renewcommand{\thesection}{S\arabic{section}}

\subsection{Holographic model and numerical methods}
For some of our numerical computations, it is more convenient to use a different coordinate system. More precisely, we consider the following ansatz for the background:
\begin{equation}
\begin{split}
\label{eq:ansatznum}
ds^2 &= \frac{1}{z^2}\left[-(1-z)\,U(z)\,e^{-\xi(z)}\,dt^2 + \frac{dz^2}{(1-z)\,U(z)} + dx^2 + dy^2\right], \qquad \phi=\phi(z),\\
A &= (1-z)\,a_t(z)\,dt - \frac{B}{2}\left(y\,dx - x\,dy\right), \qquad \chi_i = k\,x_i,
\end{split}
\end{equation}
where $z\in[0,1]$, with the AdS$_4$ boundary at $z=0$ and the black-hole horizon at $z=1$.
At the horizon, we impose regularity: $U$, $\xi$, $a_t$, and $\phi$ are finite and smooth at $z=1$, with $U(1)>0$. The factor $(1-z)$ in $A_t$ ensures that $A_t$ vanishes at the horizon. The Hawking temperature is then
\begin{equation}
T = \frac{U(1)\,e^{-\xi(1)/2}}{4\pi}.
\end{equation}
At the boundary, we impose
\begin{equation}
U \sim 1, \qquad \xi \sim 0, \qquad a_t \sim \mu, \qquad \phi \sim \phi_0\, z^{\Delta_-},
\end{equation}
where $\mu$ is the chemical potential and $\phi_0$ is the source of the dual scalar operator. The dilaton falloff follows the standard mass--dimension relation in $d=3$,
\begin{equation}
m_\phi^2 = \Delta(\Delta-3), \qquad m_\phi^2 = -V''(0) = -2,
\end{equation}
which gives $\Delta_-=1$ and $\Delta_+=2$, i.e.\ $\phi \simeq \phi_0\, z + \phi_1\, z^2 + \dots$ near the boundary. Both quantizations are allowed since $m_\phi^2$ lies in the window $-9/4 < m_\phi^2 < -5/4$. Here we take $\Delta_-$ as the source. The entropy density is obtained from the horizon area, $s=4\pi C_h$, which in the coordinates above gives $s=4\pi$ in units of the horizon radius.

To compute the local effective $\eta$ parameter, we first consider the combination
\begin{equation}
\mathcal{C} \equiv \frac{|g_{tt}|\,g_{XX}}{g_{xx}^{2}}, \qquad X \equiv \log g_{xx},
\end{equation}
where $g_{XX}=g_{rr}\,(dr/dX)^2$ is the radial metric component in the coordinate $X$.
Since $g_{tt}$, $g_{xx}$ and $X$ are scalars under radial reparametrizations $r\to\tilde r(r)$, and $g_{XX}$ is defined geometrically, $\mathcal{C}$ is invariant under such reparametrizations.
Under constant rescalings of $t$ and $x_i$, which preserve the ansatz, $\mathcal{C}$ only changes by a constant factor, so its logarithmic derivative with respect to $X$ is invariant under these as well.
This allows a direct comparison of numerical solutions written in different radial coordinates and with different normalizations of $t$, $x_i$ and $\widetilde L$.

For the $\eta$-geometry in $d+1$ bulk dimensions,
\begin{equation}
ds^2 = r^{-\frac{2\eta}{d-1}}\left[\frac{1}{r^2}\left(\frac{L^2}{f(r)}\,dr^2 - f(r)\,dt^2\right) + \widetilde L^{2}\,dx_i^2\right],
\end{equation}
one finds $g_{XX}\propto r^{-\frac{2\eta}{d-1}}/f(r)$, so the blackening factor cancels in $|g_{tt}|\,g_{XX}$ and
\begin{equation}
\mathcal{C}\;\propto\; r^{-2}\;\propto\; g_{xx}^{\frac{d-1}{\eta}} .
\end{equation}
The combination therefore encodes $\eta$ as a logarithmic slope, valid at any temperature. We use this to define a radially dependent effective exponent,
\begin{equation}
\frac{d-1}{\eta_{\rm eff}} \equiv \frac{d}{dX}\log\!\left(\frac{|g_{tt}|\,g_{XX}}{g_{xx}^{2}}\right),
\end{equation}
which is constant wherever the geometry has $\eta$-scaling.

For the ansatz \eqref{eq:ansatznum} we have $g_{xx}=z^{-2}$, hence $X=-2\log z$ and $d/dX=-\tfrac{z}{2}\,d/dz$. Moreover $g_{XX}=\big[4(1-z)U\big]^{-1}$, so $\mathcal{C}=\tfrac14\,z^{2}e^{-\xi(z)}$ and
\begin{equation}
\frac{d-1}{\eta_{\rm eff}(z)} = \frac{1}{2}\,z\,\xi'(z) - 1 .
\end{equation}
Using the different ansatz, Eq.~\eqref{eq:ansatz:gen} in the main text, this definition reduces to 
\begin{equation}
\label{eq:etaeff}
   \eta_{\rm eff}(r)\equiv 2\,(\log C)'/[\log(D\mathcal{B}/C'^{2})]'. 
\end{equation}

\subsection{Mode analysis around the intermediate scaling geometry}
We perturb the $\eta=2$ solution at $T=0$ in the gauge $\delta g_{xx}=0$,
\begin{equation}
D(r)\sim \frac{1}{r^4}\left(1+\delta d(r)\right)\,,\quad \mathcal{B}(r)\sim \frac{4}{r^4}\left(1+\delta b(r)\right)\,, \quad C(r)\sim \frac{2k^2}{3r^2}\,, \quad \phi(r)\sim 2\sqrt{2}\log(r) + \delta \phi(r)\,,
\end{equation}
Linearizing the equations of motion in $(\delta d,\delta b,\delta\phi)$ gives
\begin{equation}
\begin{split}
\frac{3}{2r^2}\left(X_B(r)+X_q(r)\right)&=\frac{\delta d''(r)}{2}-\frac{2 \delta d'(r)-\delta b'(r)}{r}-\frac{6 }{r^2}\delta b(r)-\frac{3 \sqrt{2} }{r^2}\delta \phi (r)\,,\\
-\frac{3}{2r^2}\left(X_B(r)+X_q(r)\right)&=-\frac{\delta d''(r)}{2}+\frac{\delta d'(r)-2\delta b'(r)}{r}-\frac{2\sqrt{2}}{r}\delta \phi'(r)+\frac{6 }{r^2}\delta b(r)-\frac{3 \sqrt{2} }{r^2}\delta \phi (r)\,,\\
\frac{3}{2}\left(X_B(r)+X_q(r)\right)&=3\delta b(r)+\frac{r}{2}\left(\delta d'(r)-\delta b'(r)\right)+3\sqrt{2}\delta \phi\,,\\
\frac{3r^2}{8\sqrt{2}}\left(X_B(r)-X_q(r)\right)&=\frac{1}{4}\left[r^6\frac{d}{dr}\left(\frac{\delta\phi'(r)}{r^2}+\frac{\sqrt{2}(\delta d(r) - \delta b(r))}{r^3}\right)+3\sqrt{2}\left(\delta d(r)+ \delta b(r)\right)r^2\right]+\frac{3}{2}\delta \phi(r)\,,
\end{split}
\end{equation}
where
\begin{equation}
X_B(r)\equiv \frac{\delta B^2 Z(\phi(r))}{k^2 C(r)}\,, \qquad X_q\equiv \frac{\delta q^2}{k^2Z(\phi(r))C(r)}\,.
\end{equation}
On the $\eta=2$ background, $Z=r$ and $C=2k^2/(3r^2)$, so that $X_B=3\,\delta B^2 r^3/(2k^4)$ and $X_q=3\,\delta q^2 r/(2k^4)$. Both grow toward the IR: the magnetic field and the charge density are relevant deformations of the intermediate fixed point. At the horizon, $X_q=\sigma_{\mathrm{diss}}/\sigma_{\mathrm{ccs}}=W/T$ and $X_B=W_{\Theta_H}^3/T^3$, so the two transport crossover scales of Eq.~\eqref{eq:resgen} coincide with the scales at which the corresponding deformations become of order one.
The last equation is the dilaton equation, and the third is the constraint. Substituting $(\delta d,\delta b,\delta\phi)\propto r^{n}$, we find two pairs of eigenmodes for the homogeneous case, i.e., $\delta B = \delta q=0$,
\begin{equation}
n = 0, \quad n = 3, \qquad\text{and}\qquad n = \frac{3\pm\sqrt{33}}{2},
\end{equation}
with the exponents in each pair summing to $\eta+1=3$. The mode $n=0$ is the rescaling of the time coordinate. The mode $n=3$ with $\delta d = -\delta b$ and $\delta \phi=0$ is the thermal mode, corresponding to creating a small black hole. The other pair is the dilaton-metric mode.

In the presence of small sources, $\delta B\neq 0, \delta q \neq 0$,
\begin{equation}
\begin{split}
    \delta d(r) &= -\frac{3}{4}X_B(r)\left(1-2\log r\right) - \frac{27}{16}X_q(r)\,,\\
    \delta b(r) &= -\frac{3}{4}X_B(r)\left(1+2\log r\right)-\frac{3}{16}X_q(r)\,,\\
    \delta \phi(r)&=\frac{3}{4\sqrt{2}}X_B(r)+\frac{15}{16\sqrt{2}}X_q(r)\,,
\end{split}
\end{equation}
where the logarithmic behavior is acquired since $X_B\sim r^3$ has the same radial power as the thermal mode. Note that the resonance condition $\eta+1=\eta+\gamma \kappa$ and the linear-in-temperature resistivity condition, $\gamma \kappa =1$, are equivalent in the $\eta=2$ geometry.

\end{document}